\documentclass[12pt]{article}
\usepackage{amssymb}
\begin{document}

\vskip 2cm

\begin{center}
{\bf
IMPLICATIONS OF THE SNO AND THE HOMESTAKE RESULTS FOR THE BOREXINO EXPERIMENT}

\vspace{0.3cm} S. M. Bilenky, \footnote{Also at  Joint Institute
for Nuclear Research, Dubna, Russia.} ~ T. Lachenmaier,~W. Potzel~
and  ~F. von Feilitzsch

\vspace{0.3cm} {\em Physik-Department, Technische Universit\"at
M\"unchen, James-Franck-Stra\ss e, D-85748  Garching, Germany \\}
\vspace{0.2cm}

\end{center}

\begin{abstract}
Using the recent result of the SNO solar neutrino experiment,
we have demonstrated in a model independent way that the contribution
of $^7\rm{Be}$ and other medium energy neutrinos to the event rate of the
Homestake experiment is $4 \sigma $ smaller than the
BP2000 SSM prediction. We have considered the implications of this result
for the future BOREXINO experiment.
\end{abstract}

The recently published first data of the SNO experiment \cite{SNO}
combined with the data of the Super-Kamiokande  solar neutrino
experiment \cite{S-K} present clear evidence of the presence of
$\nu_{\mu}$ and/or $\nu_{\tau}$ in the flux of high energy solar
neutrinos on the Earth. The large up-down asymmetry
of high energy atmospheric muon events, discovered by
 the Super-Kamiokande collaboration \cite{AS-K},
constitutes a model independent evidence
in favor of oscillations of atmospheric neutrinos.
 The data of all solar and atmospheric neutrino experiments
 can be described by practically decoupled two-neutrino
oscillations driven by the mixing of
three neutrinos with neutrino mass squared differences that satisfy
the hierarchy
$\Delta m^{2}_{sol} \ll \Delta m^{2}_{atm}$ (see \cite{BGG}).

There exist at present indications in favor of
 $\bar\nu_{e} \to \bar \nu_{\mu}$
oscillations that were obtained in the  accelerator experiment LSND
\cite{LSND}.
If the data of the LSND
experiment will be confirmed, we must
assume that (at least) four mixed neutrinos with small masses
exist in nature and that flavor neutrinos can transfer into
sterile states.
 A check of the LSND result is
going to be achieved by the MiniBooNE experiment \cite{MiniB},
scheduled to start in 2002.

The SNO result triggered several new global analysis of the data
of all solar neutrino experiments \cite{Bahcall,Fogli} . After the
Super-Kamiokande measurement \cite{S-K} of the energy spectrum of
recoil electrons and the zenith angle dependence of the flux of
solar neutrinos large mixing angle MSW solutions became the most
favorable fits \cite{Os} of all pre-SNO solar neutrino data
(\cite{S-K, Cl, Kam, SAGE, GALLEX}). The inclusion of the SNO data
in the global analysis only strengthens this conclusion
\cite{Bahcall,Fogli}.

The result of the SNO experiment together with results of all
other solar neutrino experiments allowed to obtain the average
values of the $\nu_{e}$ survival probability in the high energy,
medium energy and low energy regions \cite{Berez,Barger}. In
\cite{Barger} the general case of transitions of the solar
$\nu_{e}$'s into  flavor and sterile neutrinos was considered. In
\cite{Berez} the averaged values of the survival probability were
compared with the survival probability that can be inferred from
global analysis of the solar neutrino data. In  \cite{JBahcall} a
new phenomenological analysis of all solar neutrino data,
including SNO data, was done. It was assumed that there are no
transitions of solar $\nu_{e}$'s into other states and that total
solar fluxes are free fit parameters. A 7.4$\sigma $ disagreement
between the measured and the fitted values of the total event
rates was found.

The Super-Kamiokande  and SNO  solar neutrino
experiments provide evidence of oscillations of
the high energy $^8\rm{B}$ solar neutrinos. The next solar neutrino
experiment BOREXINO \cite{BOR}, scheduled for 2002, is aimed to
detect medium energy $^7\rm{Be}$, CNO and pep neutrinos.
In this paper we will infer possible implications for the
BOREXINO  experiment
 from the
results of the
SNO and the Homestake solar neutrino experiments.

The Homestake experiment
 is sensitive mainly to
 neutrinos from $^8\rm{B}$
decay and $^7\rm{Be}$ capture. According to the SSM BP2000 \cite{BP}
neutrinos from $^{15}$O and $^{13}$N decays
and the pep reaction give $\simeq 8.4$\% of the total contribution to
the chlorine event rate.

The event rate in a chlorine experiment is given by

\begin{equation}
R_{Cl} =
\int_{E_{th}}\sigma_{\nu_{e}Cl}(E)\,~\sum_{i}\Phi_{\nu_{e}}^{i}(E)\,dE\,.
\label{001}
\end{equation}

Here $\sigma_{\nu_{e}Cl}$ is the cross section of the process
\begin{equation}
\nu_{e}+^{37}\rm{Cl} \to e^{-} +^{37}\rm{Ar} \,,
\label{002}
\end{equation}

$\Phi_{\nu_{e}}^{i}(E)$ is the flux of the solar $\nu_{e}$'s from
the source $i$ on the Earth
 (i= ${^8}$B, ${^7}$Be, $^{15}$O, $^{13}$N, pep)\footnote{We will not
consider the small contribution of the hep neutrinos.} and $
E_{th}=0.814$ \,MeV is the threshold of the process (\ref{002}).
We have
\begin{equation}
\Phi_{\nu_{e}}^{i}(E)= P (\nu_{e}\to \nu_{e})\,~\Phi_{\nu_{e}}^{i0}(E)
\label{003}
\end{equation}

where $P (\nu_{e}\to \nu_{e})$ is the $\nu_{e}$ survival probability,
 $E$ is the neutrino energy and $\Phi_{\nu_{e}}^{i0}$ is the initial
flux of electron neutrinos from the source $i$.

Using now the results of the SNO experiment we will determine the
contribution of the $^8$B neutrinos to the chlorine event rate
$R_{Cl}$.

In the SNO experiment solar $\nu_{e}$'s are detected through the
observation of the charged current (CC) process
\begin{equation}
\nu_{e} + d \to e^{-}+p +p
\label{004}
\end{equation}
The neutrino energy threshold in the SNO experiment is equal to
$E_{th}=T_{th} + 1.44\, \rm{MeV} = 8.19\,\rm{ MeV}$ where
$T_{th}=6.75$ MeV is the electron kinetic energy threshold. For the
observed event rate we have \cite{Villante}
\begin{equation}
R_{SNO} = \langle \sigma_{\nu_{e}d} \rangle\, \Phi_{\nu_{e}}^{SNO}
\label{005}
\end{equation}

Here $\langle \sigma_{\nu_{e}d} \rangle $ is the cross section of
the process (\ref{004}) averaged over the initial spectrum of the
$^8$B  neutrinos and

\begin{equation}
\Phi_{\nu_{e}}^{SNO} = \langle P(\nu_{e} \to \nu_{e})\rangle _{CC}\,~
\Phi_{\nu_{e}}^{^8 B 0}
\label{006}
\end{equation}
is the flux of $\nu_{e}$ on the Earth and
$\Phi_{\nu_{e}}^{^8 B 0}$ is the total initial flux of the $^8$B
neutrinos.

In the SNO experiment\cite{SNO} it was found that
\begin{equation}
\Phi_{\nu_{e}}^{SNO} = (1.75 \pm 0.07(\rm{stat.}) \pm 0.12(\rm{sys.})
 \pm 0.05(\rm{theor.})) \cdot
10^{6}cm^{-2}\,s^{-1}
\label{007}
\end{equation}

No statistically significant
distortion of the energy spectrum of the recoil electrons in the
region 5 - 20 MeV with respect to the
spectrum, predicted under the hypothesis  of a constant survival
probability,
 was observed  in the Super-Kamiokande experiment
\cite{S-K}.
No distortion of the electron spectrum
 was observed also in the SNO experiment  in the energy
range 8.2 - 14.4 MeV.
 Thus, data of the SNO and the
Super-Kamiokande experiments are in agreement with the assumption
of a constant $\nu_{e}$ survival probability in the high energy
region.

We will assume first that the $\nu_{e}$ survival probability
for $^8$B neutrinos
is
constant in the whole energy range of the Homestake
experiment. Then we will calculate the corrections due to the
possible dependence of the survival probability on the neutrino energy in the
energy range from the threshold of the Homestake experiment
 to the threshold of the
SNO experiment.

 If we
assume a constant survival probability in the whole Homestake energy
range
we can calculate the
contribution of the $^8$B neutrinos to the chlorine event rate
using directly the SNO result. We have

 \begin{equation}
 (R_{Cl}^{^8B})_{SNO} = \int_{E_{th}}\sigma_{\nu_{e}Cl}(E)\,~
X^{^8B}(E)\,~\Phi_{\nu_{e}}^{SNO}\,dE = (2.00 \pm 0.17)\,\rm{SNU}
\label{008}
\end{equation}

Here $X ^{^8B}(E)$ is the (known) normalized spectrum of the
initial $^8$B neutrinos ($\int X^{^8B}(E)\,dE =1 $). The flux
of the $^8$B neutrinos on the Earth $\Phi_{\nu_{e}}^{SNO}$,
 measured in the SNO
experiment, is given by
 (\ref{007}).
The uncertainty in (\ref{008}) is determined by the uncertainties
of the SNO flux and of the cross section of the process (\ref{002}).

 Let us now estimate the possible effect
of the energy dependence of the probability $ P(\nu_{e} \to
\nu_{e}) $ in the energy range from the Homestake threshold
($E_{th} \simeq 0.81\, \rm{MeV}$) to the SNO threshold ($E_{th}
\simeq 8.19\, \rm{MeV}$). The contribution of the $^8$B neutrinos
to the chlorine event rate $ R_{Cl}^{^8B}$ can be presented in the
form
 \begin{equation}
 R_{Cl}^{^8B} = ( R_{Cl}^{^8 B})_{SNO} + Q
\label{009}
\end{equation}
where the quantity $ Q $ is given by

\begin{equation}
Q=\int_{0.81\,\rm{MeV}} ^{8.19\,\rm{MeV}}\,~\sigma_{\nu_{e}Cl}(E)\,
X^{^8 B}(E)\,
( P (\nu_{e}\to \nu_{e})\,~\Phi_{\nu_{e}}^{{^8B}0} - \Phi_{\nu_{e}}^{SNO})
\,~dE
\label{010}
\end{equation}

The total flux of the
active $^8$B neutrinos on the Earth, that can be inferred from the
results of the SNO and the Super-Kamiokande  experiments,
is in good
agreement with the value predicted
by the BP2000 SSM.
  Thus, we will assume that the flux
$\Phi_{\nu_{e}}^{{^8 B}0}$ in Eq.(\ref{010}) is given by the SSM.
For the survival
probability $ P (\nu_{e}\to \nu_{e})$
we used the expressions
 that can be obtained from the global analysis of the
solar neutrino data \cite{BKS}.
We have found that for the preferable large mixing angle MSW solutions
 the quantity $Q$
is significantly smaller than the error in (\ref{008}):
 $Q \simeq -0.05$ SNU for the LMA solution and $Q \simeq 0.08$ SNU
for the LOW solution.  Thus, we can neglect $Q$ in (\ref{009}) and
for the contribution  of the $^8$B neutrinos to the chlorine event
rate we can take the value (\ref{008}).

The event rate, measured in the Homestake experiment \cite{Cl}, is equal to

\begin{equation}
 R_{Cl} = (2.56 \pm 0.16 \pm 0.16) \,~~\rm{SNU}
\label{012}
\end{equation}

If we subtract (\ref{008}) from (\ref{012}) we will find a model
independent contribution of $^7$Be, CNO and pep neutrinos to the
Homestake event rate \footnote{The errors were added in
quadrature.}

\begin{equation}
 R_{Cl}^{^7 Be,CNO,pep} = (0.56 \pm 0.29) \,~\rm{SNU}
\label{013}
\end{equation}

According to the BP2000 SSM \cite{BP}, the contribution of $^7$Be neutrinos
to the chlorine event rate is equal to 1.15 SNU and the total contribution of
the medium energy neutrinos is equal to 1.79 SNU. Thus the value
(\ref{013}) constitutes a model independent 4$\sigma$ evidence,
of the solar
neutrino problem for $^7$Be and other medium energy neutrinos
based on the comparison of the SNO and Homestake results.

Let us now compare the rate (\ref{013}) with the
preferable global fits \cite{Os,Bahcall,Fogli,BKS}
of the solar neutrino data.
 In the scenario of the LMA solution the $^7$Be
contribution to the chlorine rate is equal to 0.53 SNU and the
total contribution of the medium energy neutrinos is equal to 0.77
SNU. In the case of the LOW solution the $^7$Be contribution to
the chlorine rate is equal to 0.55 SNU and the total contribution is
equal to 0.83 SNU.

Let us stress that the value (\ref{013}) was obtained directly
from the results of the SNO and the Homestake experiments. No
assumptions on the fluxes of $^7$Be, CNO and pep neutrinos and
$\nu_{e}$ survival probabilities were made. On the other side, the
global analysis of the solar neutrino data is based on the
assumption of two-neutrino oscillations and on SSM fluxes. In
spite of the large error in (\ref{013}) the fact that (\ref{013})
is compatible with the contribution of the medium energy neutrinos
to the Cl event rate, predicted by the LMA and LOW solutions, from
our point of view, is a model independent argument in favor of
these solutions.

 The SMA MSW solution
for the contribution of the medium energy neutrinos to
the chlorine event rate predicts the value 0.01 SNU, which is in 2$\sigma $
disagreement with (\ref{013}). This illustrates the preference of
the large mixing angle MSW solutions, that follows from the
direct comparison of the Homestake and SNO results.

Taking into account that (\ref{013}) is the contribution of all medium
energy neutrinos to the chlorine event rate for the contribution of
the
 prevailing 0.862 MeV
 $^7$Be neutrinos  we can obtain only an
upper bound. From (\ref{013}) we have for the  $^7$Be contribution
to the chlorine rate at a 1$\sigma$ level
\begin{equation}
 R_{Cl}^{^7Be} \leq 0.85 \,~\rm{SNU}
\label{014}
\end{equation}
This gives a
model independent upper bound for the flux of the $^7$Be neutrinos

\begin{equation}
 \Phi_{\nu_{e}}^{^7 Be} \leq 3.16 \cdot 10^{9}\,\rm{cm}^{-2}\rm{s}^{-1}\,,
\label{015}
\end{equation}

This upper bound comprises $\leq 74 $\% of the BP2000 SSM flux of the
0.862 MeV $^7$Be neutrinos.

In order to obtain from (\ref{013}) the value of the flux of the
$^7$Be neutrinos we will assume that the contributions of CNO and
pep neutrinos to the chlorine event rate $R_{Cl}^{CNO,pep}$
 are given by the
LMA and LOW MSW solutions.
We find \cite{BKS}
\begin{equation}
 R_{Cl}^{CNO,pep} = 0.24 \,~\rm{SNU} \,~(\rm{LMA});\, R_{Cl}^{CNO,pep}
 = 0.29 \,~\rm{SNU} \,~(\rm{LOW})
\label{016a}
\end{equation}

From (\ref{013}) and (\ref{016a}) we have for the contribution of
the $^7$Be
 neutrinos
to the chlorine event rate

\begin{equation}
 R_{Cl}^{^7Be} = 0.32 \pm 0.29 \,~\rm{SNU} \,~(\rm{LMA})
\label{016}
\end{equation}
and
\begin{equation}
 R_{Cl}^{^7 Be} = 0.27 \pm 0.29 \,~\rm{SNU} \,~(\rm{LOW})
\label{017}
\end{equation}
Let us stress that in (\ref{016}) and  (\ref{017}) we do not
include the $\simeq 20 $ \% estimated theoretical error \cite{BP}
of the SSM CNO flux.

For the flux of the $^7$Be neutrinos on the Earth we have
from (\ref{016}) and (\ref{017}), respectively

\begin{equation}
 \Phi_{\nu_{e}}^{^7Be} = (1.19 \pm 1.08)\cdot 10^{9}\,\rm{cm}^{-2}\rm{s}^{-1}
\label{018}
\end{equation}
and
\begin{equation}
 \Phi_{\nu_{e}}^{^7Be} = (1.00 \pm 1.08)\cdot 10^{9}\,\rm{cm}^{-2}\rm{s}^{-1}
\label{019}
\end{equation}

The fluxes (\ref{018}) and (\ref{019}) are about $3\sigma $ smaller
than the flux of the $^7$Be  neutrinos
$ 4.28\ (1.00 \pm 0.10)\cdot 10^{9}\,\rm{cm}^{-2}\rm{s}^{-1}$, predicted by
BP2000 SSM \cite{BP}.

The flux $\Phi_{\nu_{e}}^{^7 Be}$ is the product of the initial
flux of the  $^7 Be$ neutrinos $\Phi_{\nu_{e}}^{^7 Be 0}$ and the
probability of the  $^7$Be neutrinos to survive. If for the flux
$\Phi_{\nu_{e}}^{^7Be0}$ we  take the SSM value, for the survival
probability we find from (\ref{018}) and(\ref{019}) respectively
\begin{equation}
P^{^7Be}(\nu_{e} \to\nu_{e}) = 0.28 \pm 0.25
\label{020}
\end{equation}
and
\begin{equation}
P^{^7Be}(\nu_{e} \to\nu_{e}) = 0.23 \pm 0.25
\label{021}
\end{equation}

Finally we will present the rates for the future BOREXINO
experiment \cite{BOR}, expected from the phenomenological analysis
of the results of the SNO and
 Homestake experiments given above.
In the liquid scintillator  experiment BOREXINO the medium energy
$^7$Be, CNO and pep  neutrinos will be detected via the
observation of recoil electrons from elastic neutrino-electron
scattering
\begin{equation}
 \nu +e \to \nu +e
\label{022}
\end{equation}
 The energy window of the experiment is 250-800 keV.
In the calculation of the major contribution of the $^7$Be
neutrinos to the event rate we used the fluxes given by
(\ref{018}) and (\ref{019}). In the calculation of the corrections
due to CNO and pep neutrinos we used the LMA and LOW solutions.
For the expected total event rates in the BOREXINO experiment we
found, correspondingly
\begin{equation}
 R_{\rm{BOREX}} =24.4 \pm 8.6\,~ \rm{events/day};\,~
 R_{\rm{BOREX}} =22.8 \pm 8.6\,~ \rm{events/day}
\label{023}
\end{equation}
The dominant contribution to the errors in (\ref{023}) comes from
the error of the quantity $R_{Cl}^{^7Be,CNO,pep}$ given by
(\ref{013}).

It is interesting to compare the rates (\ref{023}) with the rates
predicted by the favorable large mixing angle MSW solutions. The
contribution of $^7$Be neutrinos to the BOREXINO event rate is
equal to 24.4\,events/day  for the LMA solution and 22.8\,
events/day for the LOW solution. The total contribution to the
BOREXINO event rate
 of neutrinos from all sources
is equal to 30.7\, events/day for the LMA solution and 29.0\,
events/day for the LOW solution. Thus, taking into account the
errors in (\ref{023}), the total BOREXINO event rate, predicted
directly from the results of the SNO
 and the Homestake
experiments (for the dominant  $^7$Be neutrinos), is compatible
with the rates predicted by the favorable solutions of the solar
neutrino problem.

Notice that the SMA solution for the BOREXINO event rate predicts
11.7 events/day, about 1.5$\sigma$ smaller than (\ref{022}) and
BP2000 SSM predicts 55.2 events/day, about 3.6$\sigma$ larger than
(\ref{023}).

In conclusion, from the recent result of the SNO experiment
\cite{SNO}
 and the result
of the Homestake experiment \cite{Cl} we have obtained the
contribution of  $^7$Be and other medium energy neutrinos to the
chlorine event rate. We have shown that
 this contribution is about
 $4 \sigma$ smaller than the value predicted by BP2000 SSM.
The flux of $^7$Be neutrinos we have obtained
 directly from experimental data in a model
 independent way. In the calculations of the corrections due to
CNO and pep neutrinos we used the large mixing angle MSW solutions
that after recent Super-Kamiokande \cite{S-K} and SNO  \cite{SNO}
measurements became the preferred solutions of the solar neutrino
problem. Using the flux of $^7$Be neutrinos, that was inferred
from the SNO and Homestake results, we have calculated the
expected event rate for the future BOREXINO experiment.

This work has been supported by SFB 375 f\"ur Astroteilchenphysik
der DFG. S. Bilenky acknowledges the Alexander von Humboldt
Foundation for support.

\end{document}